\documentclass[runningheads]{llncs}
\usepackage[T1]{fontenc}
\usepackage{lmodern}
\usepackage{booktabs,longtable,array,calc}
\makeatletter\@ifundefined{c@none}{}{}\makeatother 
\usepackage{amsmath,amssymb}
\usepackage{url,microtype}
\providecommand{\tightlist}{\setlength{\itemsep}{0pt}\setlength{\parskip}{0pt}}
\ifdefined{/Title (Temporal Liquidity Crossing for Blockchains: An EIP-1559 Extension for Admission Below the Base Fee) /Author (Duanyang Dan Guo)}\fi
\begin{document}
\title{Temporal Liquidity Crossing for Blockchains\\\large An EIP-1559 Extension for Admission Below the Base Fee}
\titlerunning{Temporal Liquidity Crossing for Blockchains}
\author{Duanyang Dan Guo\thanks{Contact: danguo01@gmail.com}}
\authorrunning{D. D. Guo}
\institute{TLM Research}
\maketitle
\begin{abstract}
Transactions differ in how their value changes with execution time and relative position. Ethereum's EIP-1559 {[}2{]} lets them compete through fee fields, but a fee bid gives no enforceable claim to an earlier position in the block, and no way to express willingness to accept a later one. Its base fee is a hard admission floor. We introduce \textbf{temporal liquidity markets} (TLM), a framework for crossing complementary temporal requirements: transactions seeking early-region execution (consumers) can fund transactions that opt into exclusion from that region (providers). Selected consumers complete within the early region and are ordered among themselves by authorization density, both validator-checked. In our block-local EIP-1559 extension, providers may be admitted below the base-fee floor. Under TLM regulation, funding-conditioned and capacity-capped provider gas is excluded from the ordinary EIP-1559 controller input. Counting it there instead raises the base fee and displaces ordinary demand at a positive base-fee fixed point. Exclusion lets the controller regulate ordinary demand at its original target and preserves the no-TLM ordinary allocation by construction; in synthetic experiments it raises total utilization with funded gas. Selected consumers pay their signed authorizations, which fund provider shortfalls; the surplus is burned. We state an adversary model with strategic senders and builders and prove eight properties under it, among them validator-checked rule validity, full burn of the base fee, and no subsidy. An inserted consumer bid cannot raise any other consumer's charge. With a provider fee-cap floor, a small, capped inclusion fee and a consumer allowance tied to funded provider gas by a large enough link ratio, no finite set of consumer bids or providers that the builder inserts raises its direct fee payoff.

\end{abstract}
\section{Introduction}\label{introduction}

An arbitrage expiring in one slot and a treasury rebalance due this week can use identical EIP-1559 fields while having different losses from delay. Price can correlate with urgency, but the transaction format does not identify urgency, a deadline, or willingness to accept later execution.

Time is nevertheless fundamental to execution. Transactions arrive into finite service intervals, contend for ordered state transitions, and may lose value before capacity becomes available. Current blockchains represent this information indirectly: through arrival, tips, application-level deadlines, or the user's choice of chain. Gas describes what execution consumes and fee fields describe payment limits; the temporal profile remains a secondary characteristic bundled into price and chain architecture.

\subsection{Observed workload separation}\label{observed-workload-separation}

Three broad application profiles stand out. \textbf{Latency-sensitive trading} requires rapid placement, cancellation, and relative ordering; Solana and Hyperliquid are prominent venues for this activity {[}13,16{]}. \textbf{Cost-sensitive payments and transfers} value low unit cost and can often tolerate delay; they are prominent on Tron and in the L2 ecosystem. \textbf{High-value settlement and capital-intensive DeFi} remain central to Ethereum mainnet.

The activity measures show the separation. In an August 2026 snapshot, Ethereum held \$49.76B of DeFi TVL, 8.4 times Solana's \$5.90B, while Solana recorded \$3.46B of daily DEX volume, 1.84 times Ethereum's \$1.88B {[}13,14{]}. Tron had 3.79M active addresses, 6.7 times Ethereum's 569K, but little daily DEX volume {[}13,14{]}. The temporal requirements behind these profiles differ sharply. This paper asks whether a transaction can express some of that difference directly, rather than only through fees, latency investment, or its choice of chain.

\subsection{From heterogeneous demand to mechanism scope}\label{from-heterogeneous-demand-to-mechanism-scope}

Applications can have different utility functions over execution time and position. Trading orders, liquidations, settlements, payments, oracle updates, and treasury operations can differ in sensitivity to delay, relative position, cadence, and expiry. A settlement transfer and an order-book action should not be modeled as homogeneous merely because both consume blockspace. Tokenized-asset settlement and fully on-chain order books illustrate these different settings {[}15,16{]}.

We introduce TEP and TSP as components of a framework for building \textbf{temporal liquidity markets} (TLM). A \textbf{Temporal Execution Profile} (TEP) describes one request's admissible positions, value over position, resource requirement, and expiry. A \textbf{Temporal Stream Profile} (TSP) describes cadence and continuity across blocks. They identify when a request seeking earlier service can fund a compatible request willing to execute later.

This framework can be studied on L1s and L2 sequencers. We use Ethereum because EIP-1559 supplies a mature baseline. The paper analyzes one opt-in, block-local TEP mechanism. It asks whether transactions that supply temporal liquidity can be funded and admitted without reducing the allocation of opt-out ordinary transactions, while selected temporal-liquidity consumers receive an assurance of completion within a stated early region of the block.
For request \(i\), let \(q\in A_i\) denote a feasible execution position and define its TEP as

\[
\theta_i=(A_i,V_i(q),r_i(q),d_i).
\]

Here \(A_i\) is the admissible position set, \(V_i(q)\) the value of execution at position \(q\), \(r_i(q)\) a state-dependent resource requirement, and \(d_i\) an optional expiry. The TEP is an analytical description; this paper does not propose encoding all of its fields in an Ethereum transaction.

\textbf{Definition 1 (temporal liquidity).} Temporal liquidity is the economically relevant capacity to move execution across admissible positions without violating feasibility, together with the value gained or surrendered by that movement.

\subsection{A two-sided temporal-liquidity crossing framework}\label{a-two-sided-temporal-liquidity-crossing-framework}

This subsection isolates complementary requirements for earlier and later execution before specifying EIP-1559 settlement. Consider two feasible positions \(q_E\) and \(q_L\), with \(q_E\) earlier than \(q_L\), and define transaction \(k\)'s value of moving earlier as
\[
\Delta_k=V_k(q_E)-V_k(q_L)\ge 0.
\]
Let urgent transaction \(u\) initially occupy \(q_L\), and flexible transaction \(f\) initially occupy \(q_E\). We call them a \emph{temporal-liquidity consumer} (consumer) and a \emph{temporal-liquidity provider} (provider), respectively. Let \(W_{\mathrm{before}}\) and \(W_{\mathrm{after}}\) denote their combined temporal value before and after their positions are exchanged:
\[
W_{\mathrm{before}}=V_f(q_E)+V_u(q_L),\qquad
W_{\mathrm{after}}=V_f(q_L)+V_u(q_E).
\]
The swap changes combined temporal value by \(\Delta_u-\Delta_f\). The consumer's maximum temporal bid is \(\Delta_u\); the provider's minimum temporal offer is \(\Delta_f\). A mutually acceptable crossing exists when \(\Delta_u\ge\Delta_f\); strict inequality leaves positive temporal surplus. Both are private values that the mechanism does not observe: a consumer's authorization is a bid, not a revealed \(\Delta_u\).

We apply the temporal-liquidity crossing framework to Ethereum through a \emph{Temporal Crossing Mechanism for EIP-1559}. Transactions submit bids or offers according to their temporal-liquidity roles: a consumer signs a positive temporal authorization as its bid for earlier service, while a provider's below-floor fee fields determine a shortfall, the minimum monetary offer required for accepting later service. In this framework, these correspond to the consumer's \(\Delta_u\) and provider's \(\Delta_f\). Settlement uses a pool rather than a direct transfer: selected consumers pay their bids into it, the pool covers funded provider offers, and what it does not spend is burned. Here, ``earlier'' means completion within a defined early region of the block, while ``later'' means execution outside that region. The formal rule ranks qualifying consumers by authorization per unit of realized gas while retaining builder discretion over opt-out transactions and all ordering not expressly constrained.

A full assessment additionally models block capacity, state-dependent execution costs, and effects on other transactions and protocol stakeholders.

A request set is \textbf{temporally degenerate} over a horizon when moving any request among its admissible positions changes neither its value nor what can be executed. A TEP then adds no allocative information beyond a single-resource capacity model: admission reduces to
\[
\sum_i x_i w_i\le K_{\mathrm{tot}},\qquad x_i\in\{0,1\},
\]
where \(w_i\) is request \(i\)'s fixed resource quantity, \(x_i\) its admission indicator, and \(K_{\mathrm{tot}}\) the aggregate capacity for the horizon, both in the same unit: gas in the EIP-1559 specialization. Temporal mechanisms matter when time or position changes value or feasible execution.
\subsection{Contributions}\label{contributions}

We make three contributions. (i) We introduce \emph{temporal liquidity} as an execution-time dimension orthogonal to price and execution value, and use a minimal temporal execution profile to describe it. An ordinary fee bid supplies no offer to accept later execution and no enforceable claim to position. The mechanism below supplies both. (ii) We give a two-sided temporal-liquidity crossing framework and instantiate it as a block-local extension to EIP-1559. Selected consumers' conditional authorizations fund providers that opt into execution outside the early region, enabling admission of demand otherwise suppressed by the base-fee floor. Selected consumers receive validator-checked completion within that region and relative rank among selected consumers by authorization density. The crossing is internal and self-funded: consumer payments cover provider shortfalls, the surplus is burned, and no subsidy is required from the protocol or opt-out users. (iii) We analyze TLM-adjusted base-fee metering: gas of funded providers is excluded from the controller's metered input, so temporal-liquidity information informs base-fee control. At a positive base-fee fixed point, this preserves EIP-1559's ordinary-gas target and permits total block utilization above its 50\% target. A reference-preserving admission rule adds providers only within residual capacity, retaining every transaction in the no-TLM reference block by construction rather than by measurement. We derive realized-settlement and provider-gas constraints, state an adversary model and the properties that hold under it, and show that a builder gains nothing by inserting a consumer bid or a provider of its own.

\section{Why temporal-liquidity crossing?}\label{why-temporal-liquidity-crossing}

\subsection{Limits of one-sided fee bidding}\label{limits-of-one-sided-fee-bidding}

With base fee \(b\), fee cap \(m\), and priority-fee cap \(p^{\max}\), EIP-1559 admits an ordinary transaction only if \(m\ge b\); its effective tip is \(\min\{p^{\max},m-b\}\). Ethereum's consensus rules impose no intra-block ordering: descending priority-fee order is a client default rather than a protocol rule, and competitive builder auctions need not follow it {[}24{]}. A higher tip therefore raises inclusion probability and tends to place a transaction earlier by convention, but purchases no verifiable claim to any region of the block. A transaction with \(m<b\) has neither an ordinary execution path nor a payment counterparty through which it can fund that execution. Lowering a priority-fee cap does not create an offer to execute later, while a consumer's higher payment cannot fund below-floor admission of a provider. A block-local crossing adds these missing links alongside EIP-1559's inter-block base-fee feedback.

\subsection{Block-local temporal crossing}\label{block-local-temporal-crossing}

A consumer's positive authorization is conditional: nothing is charged unless the consumer is selected, and a consumer is selected only against funded providers. Selected consumers receive completion within a defined early region; the pool covers providers' fee shortfalls in return for later execution. Validators check early-region completion and, among selected consumers, descending authorization per unit of realized gas. Opt-out transactions may still appear in the early region at builder discretion; all ordering not expressly constrained remains unchanged.

\subsection{Related work}\label{related-work}

Babaioff and Nisan analyze EIP-1559 welfare with patient bidders {[}1{]}. Their patience is cross-block waiting without a payment transfer; our providers are compensated for later in-block execution and may be admitted below the base-fee floor. Kiayias et al. propose tiered price--delay schedules for heterogeneous urgency {[}17{]}; our two-sided crossing is not a menu of service tiers or tier-level price discrimination. Work on latency races, multidimensional fees, fee-market dynamics, empirical outcomes, patient users, and delay cost provides complementary context {[}3,4,6,7,8,9,11{]}.

Mini-block auctions {[}18{]} and private-transaction evidence {[}19{]} support the premise that intra-block position can matter. Timeboost {[}12,20{]} exposes priority in production, while its concentration and reversion evidence leaves access and fairness open. Preconfirmations {[}21{]} provide advance, rather than block-local, assurance.

Roughgarden {[}10{]} evaluates transaction fee mechanisms by user incentive compatibility (UIC), myopic miner incentive compatibility (MMIC), and resistance to off-chain agreements. The miner there is the block producer, which in this paper is the builder. A first-price auction is MMIC and not UIC; second-price-type auctions fail MMIC, because the producer gains from bids it inserts. Chung and Shi {[}25{]} show that a mechanism that is UIC and proof against a producer colluding with one user must pay the producer nothing. Our consumer charge is first-price; Section~5 states what holds for inserted transactions and what is not claimed.

\section{A two-sided mechanism for EIP-1559}\label{a-two-sided-mechanism-for-eip-1559}

Let \(b\) be the block base fee. For transaction \(i\), the retained EIP-1559 fields are \texttt{max\_fee} \(=m_i\), \texttt{max\_priority\_fee} \(=p_i^{\max}\), gas limit \(L_i\), and realized charged gas \(g_i\le L_i\). It additionally signs one lump-sum \textbf{Temporal Liquidity Authorization} (\texttt{TLA}) denominated in wei. We call the signed value the transaction's \emph{declaration}; a consumer's declaration is also called its authorization or its bid. A transaction with \texttt{TLA} \(=0\) is neutral (opt-out) and follows ordinary EIP-1559 treatment. Block positions are measured by cumulative charged gas; the \emph{early region} is the prefix ending at \(E\).

\subsection{Consumers}\label{consumers}

A consumer signs \texttt{TLA} \(=c_i>0\). It is a lump-sum temporal authorization, distinct from the gas-priced EIP-1559 fee fields. Let \(g_i\) be its realized charged gas in the executed block and define its authorization density as

\[
\rho_i=\frac{c_i}{g_i}.
\]

The mechanism fixes a protected consumer-selection gas allowance \(K_C=\kappa_CK\), where \(K_C\) has gas units, \(K\) is the block gas limit, and \(\kappa_C\in[0,1]\) is the dimensionless allowance fraction. \(K_C\) caps the realized gas of consumers selected for protected service; it is not the size of the early region. Protected service is sold only against provider supply: the consumer allowance of a block is
\[A_C=\min\{K_C,\ \lambda G_P\},\]
where \(G_P\) is the realized gas of the funded providers (\S~3.4) and \(\lambda>0\) is a protocol parameter. We call this dependence the \emph{allowance link}; it binds when \(\lambda G_P<K_C\). The protocol sets \(\lambda\) so that \(K_C/\lambda\), the funded provider gas that opens the whole allowance, is at most \(\frac{\nu-\varepsilon}{\varepsilon}g_0\), with \(\nu\), \(\varepsilon\) and \(g_0\) defined in \S~3.3; this is what keeps a builder from gaining by inserting transactions (\S~5.3). A candidate is eligible if it satisfies ordinary EIP-1559 validity, including \(m_i\ge b\), appears in the proposed block, and meets the \emph{consumer floor} \(\rho_i\ge\beta b\), where \(\beta>0\) is a protocol parameter. Positive TLA does not guarantee inclusion or protected service. Under the stated selection policy, the builder processes eligible candidates by decreasing \(\rho_i\), with transaction hash breaking ties, and adds a candidate to the selected set \(C^\star\) if its realized gas fits the remaining \(A_C\); otherwise it skips the candidate. Density and the allowance use realized gas, as the provider side does (\S~3.4). The signed gas limit remains an EIP-1559 validity and liability bound and plays no part in selection, so a transaction takes allowance only by the gas it uses, on which it also pays the base fee. The allowance makes protected service scarce, so a large collection of low-density positive-TLA bids cannot all qualify. Only selected, included consumers belong to \(C^\star\) and are charged. Excluded and unselected positive-TLA transactions pay no temporal charge and remain ordinary transactions if included.

A selected consumer pays its ordinary EIP-1559 charge plus a temporal charge equal to its authorization, \(a_i=c_i\) (\S~3.4). At validity checking, the sender must back its ordinary EIP-1559 maximum liability, transferred value, and \(c_i\). A consumer therefore knows its temporal charge when it signs: \(c_i\) if it is selected, and nothing otherwise. Consumer selection is a constrained multi-unit allocation problem; this paper uses the stated deterministic greedy rule rather than the ascending-bid auction of Ausubel {[}22{]}.

\subsection{Consumer position assurance}\label{consumer-position-assurance}

TLA does not prescribe an exact transaction order. The initial mechanism uses a fixed early completion region

\[
E=\chi K
\]

with \(\chi=0.20\) and \(K_C=0.10K\) in the initial parameterization. ``Early'' is therefore a logical position, not an arrival or wall-clock time within the slot. For the final execution order \(\preceq\), let

\[
h_i=\sum_{k:\,k\preceq i}g_k
\]

be the cumulative charged-gas endpoint of transaction \(i\); its start position is \(h_i^- = h_i-g_i\). A valid block requires \(h_i\le E\) for every \(i\in C^\star\). Thus a selected consumer is fully executed before the early-region threshold is crossed. The two quantities answer different questions: \(K_C\) determines which scarce set of consumers qualifies for protected service, while \(E\) determines where that selected set must finish. The early region is not reserved solely for consumers: at the builder's discretion, a neutral transaction may appear before or among selected consumers, subject to transaction validity and state dependencies.
\subsection{Providers}\label{providers}

A provider signs \texttt{TLA} \(<0\), chooses \(m_i\) with \(\nu b\le m_i<b\), and accepts execution outside the early region, after completion of the protected consumer set. Under EIP-1559 a transaction with \(m_i<b\) has effective tip \(\min\{p_i^{\max},m_i-b\}<0\): it would pay its whole fee cap and leave nothing for a tip. The provider branch keeps that rule, as an exception to EIP-1559 validity rather than an instance of it. A provider must sign \texttt{max\_priority\_fee} \(=0\), or it is invalid. The field then plays no part, and no client has to interpret it. The fields are signed before the base fee is known; if the base fee of the including block is at most \(m_i\), the provider branch does not apply, and the transaction is an ordinary EIP-1559 transaction with zero tip. A funded provider pays its fee cap, all of it toward the base fee, and the builder is paid instead an \emph{inclusion fee} fixed by the protocol. For a provider of realized gas \(g_i\), the fee is \(\varepsilon b\) per gas, capped at the base fee of the smallest transaction:
\[
g_ip_i=\min\{\varepsilon b\,g_i,\ g_0b\},\qquad 0<\varepsilon<\nu,
\]
paid from the pool, where \(g_0=21{,}000\) is Ethereum's intrinsic gas, the least gas any transaction uses, so a provider can be as small as \(g_0\); the mechanism covers ordinary Ethereum transactions and bundles of them, each of which uses at least \(g_0\) and \(p_i\) is the fee per gas. The cap binds only above \(g_0/\varepsilon\) gas (2.1 million at \(\varepsilon=0.01\)). It ensures that no provider earns the builder more than the base fee burned by any transaction the builder inserts (\S~5.3). The provider's per-gas shortfall is
\[
s_i=b+p_i-m_i,
\]
which is \(b-m_i+\varepsilon b\) below the cap.

The \emph{fee-cap floor} makes a provider eligible only if \(\nu b\le m_i<b\), with \(0<\nu<1\). It is checked against the base fee of the including block, like the consumer floor, and it decides eligibility, not payment: a provider pays exactly \(g_im_i\). Both \(\varepsilon\) and \(\nu\) are protocol parameters, bounds that are not fitted to traffic. The inclusion fee does not depend on a provider's signed fields, so \(p_i^{\max}\) plays no part in the provider branch and the builder chooses no tip. Because every funded provider below the cap earns the builder the same fee per gas, a builder that maximizes its revenue funds as much provider gas as the pool can pay for, which favours the providers whose fee caps are closest to the base fee. Validators recompute shortfall, settlement, and builder revenue from the base fee and each provider's fee cap and gas.
For realized gas \(g_i\), the provider pays \(g_i m_i\); the pool pays \(g_i s_i\); the protocol burns \(g_i b\); and the builder receives \(g_i p_i\). Thus:

\[
g_i m_i+g_i s_i=g_i b+g_i p_i.
\]

The base fee remains fully paid and burned. What changes is the incidence: the provider's own payment may be below \(b\) because consumers cover the difference.

\subsection{Provider selection, execution, and settlement}\label{provider-selection-execution-and-settlement}

The builder executes a candidate final block before proposing it. Provider funding is not an entitlement: together with consumer selection (below), the builder chooses a funded-provider set \(\mathcal F\) from eligible negative-TLA candidates (transactions with signed negative TLA, \(\nu b\le m_i<b\), valid retained EIP-1559 fields, and a feasible placement outside \(E\)), subject to final funding, provider-gas, and early-region constraints. The protocol checks feasibility of the final set rather than a globally observable optimum. A provider below the base fee is in a block only if it is funded, so a block holds no record of candidates that are passed over, and a priority rule among providers could not be checked without a protocol-visible list of candidates. For reproducibility, our simulator scans eligible providers by increasing shortfall, with transaction identifier ties.

Let \(g_i\) be the realized gas of provider \(i\in\mathcal F\), and define
\[
S_i=g_i s_i,\qquad S=\sum_{j\in\mathcal F}S_j.
\]
Thus \(S_i\) is provider \(i\)'s funded amount and \(S\), the total shortfall, is what the pool spends. The pool's \emph{receipts} are \(R=\sum_{i\in C^\star}c_i\), the total payment of selected consumers, and the block is funding-valid only if \(S\le R\). Each block \(t\) has a provider-gas allowance \(\bar P_t\), a protocol parameter that caps total realized gas of funded providers: \(\sum_{j\in\mathcal F}g_j\le\bar P_t\). The final block must also obey Ethereum's hard gas limit. A provider's gas limit remains its normal signed EIP-1559 bound and sender-liability bound; it does not reserve TLM funding. The proposed block identifies \(C^\star\) and \(\mathcal F\). On replay, validators check signed declarations, the fee-cap floor, the consumer floor, consumer selection under \(A_C\), consumer density order and early completion, provider placement after \(E\), shortfalls, settlement, the provider-gas allowance, and hard gas validity. Verification is relative to the proposed block: validators confirm that the included set obeys the rules, not that it is the set a builder should have chosen, since transactions absent from a block leave no record and builders do not share a mempool.

\textbf{Joint selection.} The selected consumers \(C^\star\) and the funded providers \(\mathcal F\) constrain each other: the allowance \(A_C=\min\{K_C,\lambda G_P\}\) grows with funded provider gas \(G_P=\sum_{j\in\mathcal F}g_j\), and providers can be funded only up to the consumers' total payment \(R\). Neither set can be fixed before the other. Validators do not search for the pair \((C^\star,\mathcal F)\). They check that the pair in the proposed block is consistent: \(C^\star\) is what the selection rule of \S~3.1 yields under \(A_C\), and \(\mathcal F\) satisfies the funding and provider-gas constraints with that \(R\). A builder can find a consistent pair by starting from all candidates and only removing: select consumers under the current allowance, compute \(R\), drop providers that \(R\) cannot fund, recompute \(A_C\), and repeat, never restoring a candidate it has removed. With realized gas held fixed, each pass can only lower the allowance, so the procedure ends; its extension to realized gas that changes with the order is left to future work (\S~7). A consistent pair always exists: with no funded provider, \(G_P=0\), no consumer is selected or charged, and the block is an ordinary EIP-1559 block. The rule is therefore a validity predicate on the executed block, not an auction run before execution: the block is valid only if its realized gas, selected sets, settlement and order agree on replay.

\textbf{Settlement} is pay-as-bid. Every selected consumer pays its signed authorization, and providers are funded from the total:
\[
a_i=c_i\quad(i\in C^\star),\qquad R=\sum_{i\in C^\star}a_i,\qquad S=\sum_{j\in\mathcal F}S_j\le R.
\]
The surplus \(R-S\) is burned. No balance is carried between blocks and no external funding enters, so
\[
\sum_{i\in C^\star}a_i=R=S+(R-S).
\]
Burning is the destination that no participant can turn to its own account within the block. Returned to consumers in proportion to their authorizations, the surplus reproduces the pro-rata rule set aside in \S~3.6. Paid to the builder, it would reward higher bids. Paid to providers beyond their shortfalls, it would reward transactions that exist to collect it. Carrying the surplus to later blocks as a reserve is a multi-block design outside this paper. Pro-rata cost sharing has a related literature, including robust double auctions and excludable-public-good mechanisms {[}5,23{]}.
\subsection{Ordering}\label{ordering}

Provider selection and logical execution order are distinct. The builder may use shortfall, expected revenue, or state dependencies when choosing providers; the protocol checks only the final set's realized constraints. To check consumer order, delete every transaction other than those in \(C^\star\) from the final block order. The remaining list must be sorted by nonincreasing authorization density \(\rho_i\), with transaction hash breaking ties. Every funded provider \(j\) must begin outside the early region, \(h_j^-\ge E\). Neutral transactions may appear before or among consumers, provided \(h_i\le E\) for all \(i\in C^\star\). The mechanism constrains three things: completion of \(C^\star\) within \(E\), the density order within \(C^\star\), and the start position of funded providers. Block composition, provider order, and all else remain at builder discretion, as under EIP-1559 and proposer-builder separation (PBS) today.
\subsection{Declaration consequences}\label{declaration-consequences}

A positive declaration \(c_i\) is a sender-backed liability and, through \(\rho_i=c_i/g_i\), the input to consumer selection and to relative order within the consumer subsequence. If the consumer is selected, its temporal charge is its declaration, \(a_i=c_i\). Declaring more improves selection rank and position among selected consumers, and costs the amount declared. A declaration below the consumer floor \(\beta b g_i\) does not qualify, and when consumer demand exceeds the allowance, low-density candidates are skipped.

\textbf{The rule is first-price.} A consumer's charge is set by its own declaration and by no other transaction. It is therefore not UIC in the sense of {[}10{]}: a consumer gains by declaring less than its value, as far as it can and still be selected, and its best declaration depends on declarations it does not see. Where declarations are visible before the block, rivals can also outbid one another in steps {[}26{]}. A declaration above value does not pay, since it is charged in full. EIP-1559's priority fee is also paid as bid, but it buys no enforceable position, and under EIP-1559 the posted base fee does the rationing when blocks are not full {[}10{]}; the temporal charge has no posted price beyond the consumer floor. Section~6 reports how funded provider gas varies with the level of declarations.

\textbf{Two alternatives are set aside.} (i) \emph{Pro rata.} Charging \(a_i=c_iS/R\), with \(R=\sum_{i\in C^\star}c_i\) the total authorized, and refunding \(R-S\) is budget-balanced, but it bounds every charge by the funded shortfall \(S\), an amount that providers determine. A consumer that values rank above \(S\) gains by inflating its declaration, limited only by the balance it can back, and when \(S=0\) protected service is free. (ii) \emph{Generalized second price.} Charging each selected consumer the next density in the selection order, so that consumers pay different prices by rank, removes that incentive, but lets one transaction's declaration set another's charge. A builder earns the inclusion fees of the providers it funds, so it gains by inserting a bid just below a selected consumer (\S~5); second-price rules are the standard example of this failure {[}10{]}. Unlike a second-price auction for one item, this rule is not truthful: a consumer that drops below the next one pays a lower density, as in generalized second-price position auctions. In our traffic model it leaves 94\% of selected consumers able to gain by declaring less, and it funds the same provider gas as pay-as-bid once each consumer pays the least at which it is still selected (\S~6).

A negative declaration opts the transaction into the provider branch: it executes outside \(E\), and funded admission requires a fee cap below the base fee. The provider pays \(g_i m_i\); the pool pays the shortfall needed for the burn and the inclusion fee.

\textbf{Lemma 1 (negative-TLA magnitude is inactive).} \emph{Holding the fee fields, realized gas, and other block inputs fixed, a provider's negative TLA value affects neither its shortfall, settlement, funding feasibility, nor ordering requirement.}

\emph{Proof.} Only the sign selects the provider branch; shortfall and settlement are defined from the fee fields and realized gas, and placement is after \(E\).

The mechanism is block-local. It reallocates position and payment inside one block; it does not defer a transaction across blocks or create physical capacity.

\section{Evaluating temporal liquidity in a blockchain economy}\label{evaluating-temporal-liquidity-in-a-blockchain-economy}

\subsection{Evaluation framework}\label{evaluation-framework}

The temporal-liquidity framework models a blockchain economy as a mix of applications whose demand differs in urgency, delay tolerance, expiry, and position value. Traffic mix is therefore an economic input, not merely a simulation detail: it determines whether complementary temporal demand and supply can cross. Capacity alone is not a sufficient benchmark, because more admission can displace other demand or change incidence. We report settlement, admission and displacement, controller response, and incidence. The reference-preserving admission rule, defined in \S~4.4, retains the no-TLM reference block; consumer charges are the signed authorizations of selected consumers.

The synthetic traffic model has positive-TLA consumers, below-floor providers, and neutral transactions. Consumers seek early-region service and sign heterogeneous authorizations, so that selection and rank vary with willingness to pay for earlier execution; providers accept execution outside that region; neutral transactions retain the baseline path. These are economic roles, rather than fixed application labels. Homogeneous traffic or traffic without providers reduces to the EIP-1559 baseline. Let their offered-gas shares be

\[
(\pi_C,\pi_P,\pi_N),\qquad \pi_C+\pi_P+\pi_N=1.
\]

Our simulator compares baseline EIP-1559 with two TLM variants. The main offered-gas mix is \((\pi_C,\pi_P,\pi_N)=(1/5,1/3,7/15)\). Additional sensitivity runs set \(\pi_P=1/3\), vary \(\pi_C\in\{1/10,1/5,1/3\}\), and assign the residual share to \(\pi_N\). In the \(\pi_C=1/3\) run, the offered gas of positive-TLA consumer candidates exceeds the protected consumer-selection allowance \(K_C=0.10K\), so the allowance is active and candidates are skipped. Runs contain 600 blocks (200 warm-up) over 20 fixed seeds. The hard limit is \(K=60\) million gas, the limit Ethereum adopted in November 2025, and the EIP-1559 target is \(g^\ast=30\) million gas. Provider-gas allowances are \(\bar P_t\in\{0.20K,0.30K\}\), and the early-region capacity is \(E=0.20K\); each is measured in gas. The consumer floor is \(\beta=0.10\), the inclusion fee is \(\varepsilon=0.01\), the fee-cap floor is \(\nu=1/2\), and the link ratio is \(\lambda=6\), so the full consumer allowance needs \(K_C/\lambda=1\) million gas of funded providers, below \(\frac{\nu-\varepsilon}{\varepsilon}g_0=1.03\) million. A builder's own provider costs it at least \(0.49b\) per gas. The fee-cap floor does not bind for the simulated fee caps. The cap on the inclusion fee binds only above 2.1 million gas, and simulated provider gas is at most 1.5 million. Ordinary and consumer fee caps follow a capped Pareto distribution with tail exponent 1.35, and consumer authorizations are drawn as realized gas times a uniform factor on \([4,12]\). Declarations are thus drawn, not chosen: the simulation does not model bidding, and Section~6 reports sensitivity to their level. Because simulated gas is independent of fee caps, this exponent is also the base-fee elasticity of ordinary gas demand; other inputs are synthetic. Results report 95\% Student-\(t\) intervals. The simulator checks selection, settlement, early-region completion, admission, and the gas measure used by the base-fee controller.
\subsection{TLM regulation for the base fee}\label{tlm-regulation-for-the-base-fee}

Let \(O_t\) denote controller load from neutral and consumer transactions (not only TLA-zero transactions), and let \(P_t\) denote funded provider gas. We compare \emph{na\"ive metering} with \emph{TLM regulation}. Under na\"ive metering, the EIP-1559 controller measures \(G_t=O_t+P_t\), so provider gas is included in the controller load. Under TLM regulation, the controller measures \(G_t=O_t\). Both use the EIP-1559 base-fee feedback rule {[}2,6,10{]}.
\[
b_{t+1}=b_t\left[1+\gamma\frac{G_t-T}{T}\right],
\]
where \(T=g^\ast\) is the EIP-1559 target and \(\gamma>0\) is the adjustment parameter. A base-fee fixed point means \(b_{t+1}=b_t>0\): the controller leaves the base fee unchanged from one block to the next. Provider gas is excluded because it is not admitted through the same ordinary EIP-1559 base-fee regulation as \(O_t\): its admission is conditional on temporal-liquidity funding from consumers and bounded by the TLM provider allowance. The TLM controller therefore treats \(O_t\) as ordinary congestion and \(P_t\) as a separately funded, capacity-capped addition.

\textbf{Proposition 1 (fixed-point identity).} \emph{Under na\"ive metering, if the base fee is at a positive fixed point, then \(O_t+P_t=T\). Thus, funded provider gas uses part of the metered target that would otherwise be available to other ordinary transactions.}

\emph{Proof.} At a fixed point, divide the update by \(b_t\) to obtain \(1=1+\gamma(G_t-T)/T\). Since \(\gamma>0\) and \(T>0\), \(G_t=T\). Under na\"ive metering, \(G_t=O_t+P_t\).

Proposition~1 is the negative half of the comparison: when all execution is included in the controller load, funded provider gas cannot raise fixed-point total utilization above \(T/K\); it replaces other metered gas. TLM regulation is designed to retain the ordinary controller target while admitting bounded, funded provider gas in addition.

In the main mixture, na\"ive metering leaves utilization at the baseline level, 50.07\% against 50.05\%: it admits 6.25 million provider gas per block, but ordinary admitted gas falls from 30.03 to 23.79 million and the mean base fee rises from 28.54 to 33.78. Metered gas stays at target while funded provider gas displaces ordinary admission. Under TLM regulation with 20\% and 30\% provider allowances, ordinary gas remains 30.03 million at a 28.54 base fee and total utilization reaches 69.33\% and 71.57\%.

\subsection{Base-fee increase under na\"ive metering}\label{base-fee-increase-under-naive-metering}

\textbf{Proposition 2.} \emph{Let \(D(b)\) be ordinary gas demand at base fee \(b\), \(b_0\) the baseline fixed-point base fee with no funded provider gas, and \(b_1\) the na\"ive-metering fixed-point fee with funded provider gas \(P^\ast=\phi g^\ast\). Suppose \(0\le\phi<1\), \(\eta>0\) is the absolute base-fee elasticity of \(D\), \(D(b_1)/D(b_0)=(b_1/b_0)^{-\eta}\), baseline demand clears at the EIP-1559 target \(D(b_0)=g^\ast\), and the hard limit does not bind. Then:}

\[
\frac{b_1}{b_0}=(1-\phi)^{-1/\eta}.
\]

\emph{Proof.} At the new fixed point, ordinary demand is \(D(b_1)=g^\ast-P^\ast=g^\ast(1-\phi)\). Constant-elasticity demand gives \(D(b_1)/D(b_0)=(b_1/b_0)^{-\eta}\). Equating the expressions gives the result.

The point is to quantify the price of na\"ive metering: provider gas \(P^\ast\) forces ordinary gas down to \(g^\ast-P^\ast\), so the base fee rises to ration ordinary demand. In the 20\%-consumer run, \(P^\ast/g^\ast=0.208\); Proposition~2 predicts \(b_1/b_0=1.189\) under the synthetic Pareto fee-cap exponent \(\eta=1.35\), versus 1.183 in simulation. The remaining 0.5\% difference reflects the capped Pareto generator and indivisible transactions, whereas Proposition~2 assumes an uncapped constant-elasticity demand curve.

\subsection{A bounded TLA-funded provider allowance}\label{a-protected-temporal-allowance}

Consumer funding creates financial capacity, not physical gas capacity. Let \(K\) be the hard block gas limit and \(T\) the EIP-1559 gas target used by the ordinary congestion controller, with \(T<K\); \(O_t\) denotes neutral and consumer gas, and \(P_t\) funded provider gas. Keeping the base fee unchanged does not itself preserve ordinary transactions: providers can still displace them within \(K\).

Consumers are selected under the allowance rule of \S~3.1, giving total consumer payment \(R=\sum_{i\in C^\star}c_i\) in wei. We then use a \emph{reference-preserving admission rule} with two stages. Stage 1 constructs a reference ordinary block \(B_t^0\), the block that would have been built from neutral transactions and consumers without TLM, with realized gas \(H_t^0\); our simulator packs it greedily by effective EIP-1559 priority fee, ties broken by a fixed identifier, so the same input always yields the same reference block. Stage 2 adds providers without removing any transaction in \(B_t^0\); every consumer candidate that the no-TLM reference policy includes stays in \(B_t^0\) whether or not it is selected, so the set of ordinary transactions does not depend on selection. The builder executes the final block; provider admission then has exactly two realized constraints:

\[
\sum_{j\in\mathcal F_t}g_js_j\le R,
\qquad
\sum_{j\in\mathcal F_t}g_j\le A_t,\quad A_t=\min\!\left\{\bar P_t,K-H_t^0\right\}.
\]

The provider-gas allowance \(\bar P_t\) is defined in \S~3.4; \(A_t\) is the residual provider-admission budget that additionally preserves the reference block. The first constraint says selected consumers fund actual provider shortfalls; \(K_C\) bounds this funding side by limiting who can be selected. The second caps actual provider gas and preserves the reference block. Both constraints are evaluated on realized gas, so no pool funding is committed before execution.

\textbf{Lemma 2 (non-displacing provider admission).} \emph{Under the reference-preserving admission rule, every transaction in \(B_t^0\) is retained and providers are added only within residual capacity. The completed block is therefore physically feasible, and realized gas increases by funded realized provider gas (equivalently, utilization increases by \(P_t/K\)).}

\emph{Proof.} The first stage is retained. The realized provider-gas constraint ensures that provider gas fits within \(K-H_t^0\). Since the ordinary allocation is unchanged, the increase in realized gas is funded provider gas.

Under TLM regulation, provider gas is excluded from the ordinary congestion signal, so the controller sees only \(O_t\):

\[
b_{t+1}=b_t\left[1+\gamma\frac{O_t-T}{T}\right].
\]

At a positive fixed point of the TLM controller, the same calculation as Proposition~1 gives \(O^\ast=T\). Together with Lemma~2, any funded \(P^\ast>0\) gives

\[
U^\ast=\frac{T+P^\ast}{K}>\frac{T}{K}.
\]

For \(T=K/2\), TLM regulation therefore raises total utilization above 50\% while preserving the ordinary base-fee controller target and the reference ordinary allocation. With the same ordinary demand curve and target, TLM regulation retains the baseline fixed-point base fee; providers remain subject to the hard block limit and provider allowance.

\subsection{Heterogeneous preferences and Pareto comparison}\label{heterogeneous-preferences-and-pareto-evaluation}

Let \(V_i(q_i)\) be request \(i\)'s gross value, in utility units, when it completes at logical block position \(q_i\), and let \(z_i\) be its total monetary payment. We use the simple separable utility model

\[
U_i(q_i,z_i)=V_i(q_i)-\mu_i z_i,\qquad z_i\le B_i.
\]

Here \(\mu_i>0\) converts money into utility loss (payment sensitivity), and \(B_i\) is a budget. These are behavioral variables, not transaction fields. Let \(q_i^0\) denote the consumer's position in the no-TLM reference block and \(q_i^1\) its realized position under TLM. A selected consumer is individually rational only if its temporal value gain covers its realized TLA charge:

\[
V_i(q_i^1)-V_i(q_i^0)\ge\mu_i a_i.
\]

A provider has several outside options: raise its fee cap to \(m_i\ge b\) and seek ordinary EIP-1559 admission, defer or cancel, or use another venue. Let \(U_i^{\mathrm{out}}\) denote the best such utility. Funded provider participation is individually rational only if its TLM utility is at least \(U_i^{\mathrm{out}}\); the pool payment covers the provider's shortfall and is not cash received by that provider.
The reference-preserving rule makes the incidence by party explicit:
\begin{itemize}
\item \textbf{Reference-block neutral senders:} no funded provider displaces them.
\item \textbf{Consumers:} selected consumers receive early-region completion assurance and relative rank among selected consumers by authorization density, and pay their signed authorization.
\item \textbf{Providers:} funded providers can execute below the base-fee floor, outside the early region, if this dominates their outside option.
\item \textbf{Builders:} for each funded provider \(j\), the builder receives the inclusion fee \(g_jp_j\le\varepsilon b\,g_j\) without replacing a reference transaction's fee.
\item \textbf{Protocol and ETH holders:} each funded provider still burns \(g_jb\). Relative to the reference block, this adds ETH burn and reduces ETH supply, holding other effects fixed; the surplus of consumer payments over funded shortfalls is burned as well.
\end{itemize}
Two of these hold by construction rather than by measurement: retention follows from Lemma~2, and each side's payment is bounded by its own signed declaration, \(a_i=c_i\) for consumers and \(m_i\) per gas for providers. Participation is then individually rational for any sender that authorizes no more than the movement is worth to it. Suppose both individual-rationality conditions hold, the builder's inclusion fees cover its added cost of building the crossing, and ETH holders gain from the added burn, holding price and other effects fixed. Then the rule is a weak Pareto improvement on the reference block for the consumers, providers and builder that take part and for ETH holders, and capacity is respected. Neutral senders have a separate guarantee, by construction: each neutral transaction of the reference block is kept, with its EIP-1559 fee treatment (Lemma~2). That is all EIP-1559 promises it. It holds no claim to a region of the block: its place is at the builder's discretion with or without TLM, as it is today when a builder accepts private order flow, and it may still appear in the early region. A neutral sender whose value depends on its position may gain or lose from where it lands, as it may today; a weak Pareto claim that includes it needs \(V_i(q_i^1)-\mu_iz_i^1\ge V_i(q_i^0)-\mu_iz_i^0\). Our simulator assigns no numerical \(V_i\), \(\mu_i\), budgets, outside options, or builder operating costs; its results are allocation and accounting results.

\section{Security model and incentive properties}\label{incentives-and-self-supply-safety}

For a transaction fee mechanism, security means that the rules hold against parties who act in their own interest, and that no party can gain at others' expense by means the rules do not anticipate {[}10,25{]}. This section states the parties and what each can do, lists the properties the mechanism has, with the result that establishes each, and lists what it does not have.

\subsection{Adversary model}\label{adversary-model}

\begin{itemize}
\tightlist
\item \textbf{Senders} sign the EIP-1559 fields and a declaration. A sender may choose any declaration, may send many transactions, and may be affiliated with the builder. It cannot forge another sender's signature, and its balance must back its liabilities (\S~3.1).
\item \textbf{The builder} sees the candidates it receives and chooses the included set, the order, the selected consumers \(C^\star\), and the funded providers \(\mathcal F\). It may insert transactions it controls, as a consumer or as a provider, and may omit any candidate. It maximizes its payoff in the current block, as in the myopic model of {[}10{]}. We count its \emph{direct fee payoff}: the priority and inclusion fees it receives, less the payments and burns of the transactions it controls. This excludes the execution value of those transactions, MEV, payments outside the protocol and effects on later blocks. A builder that sends a transaction for its own execution value is a sender like any other; the results of \S~5.3 concern transactions inserted to change the crossing.
\item \textbf{The proposer} chooses a builder's block. A proposer that builds its own block is a builder, and a proposer that colludes with a builder is treated as one party with it.
\item \textbf{Validators} replay the proposed block and accept it only if the checks of \S~3.4 pass. We assume, as Ethereum does, that a majority of them follow the protocol.
\end{itemize}
Outside the model are strategies across blocks, agreements made outside the protocol, attacks on consensus, and censorship at the network layer.

\subsection{Properties}\label{security-properties}

Each property below is stated against the adversary model of \S~5.1 and proved from the rules of \S~3 and the results of \S~4 and \S~5.3.

\textbf{Property 1 (rule validity).} \emph{In every accepted block, the selected consumers complete within \(E\) in order of nonincreasing density, every funded provider starts at or after \(E\), the total provider shortfall \(S\) is at most the pool's receipts \(R\) from selected consumers (\S~3.4), funded provider gas is at most \(\bar P_t\), every provider satisfies the fee-cap floor, and the block is within the hard gas limit. This holds whatever the builder includes, omits or inserts.}

\emph{Proof.} Each condition is a predicate on data the proposed block contains: signed fields, the base fee, the identified sets \(C^\star\) and \(\mathcal F\), the execution order, and realized gas obtained by executing the block. Validators recompute every predicate on replay (\S~3.4) and reject the block if one fails. Under the honest-majority assumption, a block that fails a check is not accepted. \(\square\)

\textbf{Property 2 (base-fee integrity).} \emph{The base fee \(g_ib\) of every included transaction is burned in full.}

\emph{Proof.} A neutral or consumer transaction satisfies \(m_i\ge b\) and pays its ordinary EIP-1559 charge, of which \(g_ib\) is burned. A transaction with \(m_i<b\) is valid only as a funded provider (\S~3.4). It pays \(g_im_i\) and the pool pays \(g_is_i=g_i(b+p_i-m_i)\); their sum is \(g_ib+g_ip_i\), of which \(g_ib\) is burned and \(g_ip_i\) goes to the builder. \(\square\)

\textbf{Property 3 (no subsidy).} \emph{Provider shortfalls are paid only from the temporal charges of the block's selected consumers. No balance enters from outside the block or is carried between blocks, and \(R=S+(R-S)\) with \(R-S\ge0\) burned.}

\emph{Proof.} The only payments into the pool are the charges \(a_i=c_i\) of \(i\in C^\star\), summing to \(R\). The only payments out are the shortfalls, summing to \(S\), and funding validity requires \(S\le R\) (\S~3.4). The rule burns \(R-S\) and defines no other source or destination. \(\square\)

\textbf{Property 4 (bounded signed liability).} \emph{A consumer pays its ordinary EIP-1559 charge plus a temporal charge of \(c_i\) if selected and nothing otherwise. A provider pays \(g_im_i\), at most its signed fee cap per gas. A neutral transaction is charged as under EIP-1559. Hence a consumer that declares no more than the service is worth to it is no worse off for participating.}

\emph{Proof.} The charges follow from \S~3.1, \S~3.3 and \S~3.4, and a neutral transaction never enters either branch. If a consumer with value \(V_i\) for the service declares \(c_i\le V_i\), its gain is \(V_i-c_i\ge0\) when selected and zero otherwise. A provider pays at most its signed fee cap per gas, as under EIP-1559; whether that leaves it better off depends on its outside options (\S~4.5). \(\square\)

\textbf{Property 5 (charge independence).} \emph{No other transaction, including one the builder inserts, changes the amount a selected consumer pays. Another transaction can change only whether the consumer is selected.}

\emph{Proof.} The temporal charge of \(i\) is \(c_i\) if \(i\in C^\star\) and zero otherwise; no term depends on another declaration. An inserted bid of higher density can take allowance and leave \(i\) unselected, in which case \(i\) pays nothing. This is Proposition~4(a). \(\square\)

\textbf{Property 6 (self-supply safety).} \emph{A builder loses at least \((\nu-\varepsilon)b\,g_i\) on every provider it inserts, and no finite set of inserted providers raises its direct fee payoff through the consumer allowance.} \emph{Proof.} Proposition~3, Proposition~6 and Corollary~1. \(\square\)

\textbf{Property 7 (no gain from inserted consumer bids).} \emph{No finite set of inserted consumer bids raises the builder's direct fee payoff.} \emph{Proof.} Proposition~6 and Corollary~1. \(\square\)

\textbf{Property 8 (protection of ordinary transactions).} \emph{Under the reference-preserving admission rule, every transaction of the no-TLM reference block is retained, and at a positive fixed point of TLM metering ordinary gas equals the target \(T\).} \emph{Proof.} Retention is Lemma~2. Metering excludes funded provider gas, so the argument of Proposition~1 gives \(O^\ast=T\) (\S~4.4). \(\square\)

\subsection{Self-supply and inserted bids}\label{self-supply-and-inserted-bids}

\textbf{Self-supply.} MMIC asks that a builder cannot raise its payoff in the current block by inserting a transaction it controls. We check this for the direct fee payoff of \S~5.1. A provider needs this check, because the pool pays its shortfall: a builder could insert a provider of its own, pay its fee \(g_im_i\) as the sender, and collect the inclusion fee \(g_ip_i\) as the builder. The insertion loses only if the builder receives no more than the sender pays, which we call the \emph{sender-backing condition}.

\textbf{Proposition 3 (direct self-supply safety).} \emph{The mechanism in \S~3 satisfies the sender-backing condition: a funded provider pays \(m_i\) per gas, the builder receives the inclusion fee \(p_i\le\varepsilon b\), and builder net on a self-supplied provider is \(g_i(p_i-m_i)\le-(\nu-\varepsilon)\,b\,g_i<0\), by the fee-cap floor \(m_i\ge\nu b\) and \(\varepsilon<\nu\).}

The protection comes from the sender-paid fee cap, the fee-cap floor, and an inclusion fee below it; without the floor the loss could be made arbitrarily small. The fee is fixed by the protocol, so the bound holds whatever the builder does.

\textbf{Inserted consumer bids.} A builder earns the inclusion fees of the providers it funds, and consumer payments decide how many are funded. Under a rule in which one bid sets another's charge, that is a reason to insert bids. Under pay-as-bid it is not.

\textbf{Proposition 4 (inserted consumer bids).} \emph{Let the builder insert a consumer transaction with declaration \(\hat c\) and realized gas \(\hat g\) into a block in which, without it, selected consumers pay \(R_0\), the funded providers receive \(S_0\), and \(R_0-S_0\ge0\) is burned. (a) With or without the insertion, every other consumer is charged its own declaration if selected and nothing otherwise. (b) Suppose the other selected consumers are unchanged and the builder funds an additional set \(D\) of providers. Its payoff changes by}
\[\Delta=\sum_{j\in D}g_jp_j-\hat c-b\,\hat g\ \le\ (R_0-S_0)-\sum_{j\in D}g_j(b-m_j)-b\,\hat g.\]
\emph{Hence the insertion loses whenever \(R_0=S_0\), and no insertion gains more than the payments that would otherwise be burned.}

\emph{Proof.} (a) holds by the settlement rule of \S~3.4. For (b), the inserted transaction pays \(\hat c\) as a consumer and burns \(b\,\hat g\); its own tip returns to the builder. The funding constraint gives \(\sum_{j\in D}g_js_j\le \hat c+(R_0-S_0)\), and \(g_jp_j=g_js_j-g_j(b-m_j)\) with \(m_j<b\).

Funded from its own payment, an inserted bid therefore cannot pay for itself: a provider's inclusion fee is always below its shortfall. It must also meet the consumer floor and fit the allowance like any other bid, and when the allowance is full it displaces a consumer whose payment then leaves the pool. The case that remains uses payments already made. A builder funds, without inserting anything, every provider that the residual covers, so an inserted bid is needed only when \(R_0-S_0<g_js_j\) for each \(j\in D\). Then, by (b),
\[\Delta\le(R_0-S_0)-\sum_{j\in D}g_j(b-m_j)-b\hat g<g_jp_j-b\hat g\le g_0b-b\hat g\le0,\]
since the inclusion fee is capped at \(g_0b\) and every transaction uses at least \(g_0\) gas, \(\hat g\ge g_0\). An inserted bid therefore cannot gain from the block's unspent payments, whatever the size of the provider. If the inserted bid displaces no selected consumer, every consumer then pays what it would have paid and one more provider executes; the residual, which would have been burned, pays that provider's shortfall instead. The protocol cannot tell a builder-affiliated consumer from any other, and the bid meets the same conditions and pays the same charge either way. When the allowance link binds, that is, when the consumer allowance \(A_C=\lambda G_P\) is set by funded provider gas rather than by the cap \(K_C\) (\S~3.1), an insertion can also act through the allowance; the rest of this section treats that case.

\textbf{The allowance link.} While the link binds, a funded provider of gas \(g_j\) also adds \(\lambda g_j\) to the consumer allowance, and the consumers it admits pay toward its shortfall. A builder can then insert a provider of its own, a \emph{seed}, with a fee cap near \(b\). As the seed's sender, the builder pays most of the seed's base fee itself, so the seed takes little from the pool, yet it adds allowance like any provider.

A constructed block shows the effect. Take \(\lambda=1\), which breaks the condition of \S~3.1. The only waiting candidates are consumers with 2.7 million gas in total, each paying \(0.405b\) per gas, and providers with 2.7 million gas in total, each with shortfall \(0.408b\) per gas. Funding providers of gas \(G\) admits \(G\) gas of consumers, who pay \(0.405bG\) against shortfalls of \(0.408bG\). So no set of providers can be funded, and nothing crosses.

Now let the builder insert a 21{,}000-gas seed with fee cap \(0.99b\). Its shortfall is \(b-0.99b+\varepsilon b=0.02b\) per gas. The seed and 2{,}679{,}000 gas of the providers open an allowance of 2.7 million gas, which admits every consumer. The consumers pay \(1{,}093{,}500b\), and the shortfalls total \(0.02b\times21{,}000+0.408b\times2{,}679{,}000=1{,}093{,}452b\), so the block is valid. The builder pays \(0.99b\times21{,}000=20{,}790b\) as the seed's sender and earns \(\varepsilon b\times2{,}700{,}000=27{,}000b\) in inclusion fees, a gain of \(6{,}210b\). The gain depends on the near-tie: with shortfalls of \(0.409b\), the seed could fund at most 2.02 million gas of the providers, and the builder would lose. The next two propositions state what such a gain needs and show that the condition of \S~3.1 rules it out.

\textbf{Proposition 5 (a condition for a gain).} \emph{Let the builder insert one transaction of its own, of gas \(\hat g\): either a consumer bid with declaration \(\hat c\) or a provider. Let \(I\) be the builder's block that contains the inserted transaction, \(B^\ast\) its best block among those with no inserted transaction, and \(D\) the providers, other than an inserted one, funded in \(I\) but not in \(B^\ast\); \(G_D\) is their total gas. If \(I\) pays the builder more than \(B^\ast\), the inclusion fees on \(D\) exceed the cost of the inserted transaction, so, for an inserted provider and for an inserted consumer bid respectively,}
\[G_D=\sum_{j\in D}g_j>\frac{\nu-\varepsilon}{\varepsilon}\,\hat g,\qquad G_D>\frac{\hat c+b\hat g}{\varepsilon b}\ge\frac{1+\beta}{\varepsilon}\,\hat g.\]

\emph{Proof.} Take the ordinary transactions as the same in \(I\) and \(B^\ast\). The builder's payoff from the crossing is then its inclusion fees less what it pays for its own inserted transaction: consumer charges go to the pool, not to the builder, and a consumer pays the builder the same priority fee whether or not it is selected; selection adds only its temporal charge, which goes to the pool. Compared with \(B^\ast\), \(I\) gains the inclusion fees on \(D\), loses those on the set \(L\) of providers of \(B^\ast\) that \(I\) leaves out, and pays the cost of the inserted transaction:
\[\text{gain}\ =\ \sum_{j\in D}g_jp_j\ -\ \sum_{j\in L}g_jp_j\ -\ \text{cost of the inserted transaction}.\]
The middle term is not negative, so the gain is at most the fees on \(D\) less the cost.
Since \(g_jp_j=\min\{\varepsilon b\,g_j,\,g_0b\}\le\varepsilon b\,g_j\), the fees on \(D\) total at most \(\varepsilon b\,G_D\). An inserted provider costs the builder \(\hat g(m-p)\ge(\nu-\varepsilon)b\hat g\), its fee cap less its own inclusion fee (Proposition~3). An inserted consumer bid costs it \(\hat c+b\hat g\ge(1+\beta)b\hat g\), its declaration and the base fee it burns; its tip returns to the builder. If the insertion gains, then
\[0\ <\ \text{gain}\ \le\ \varepsilon b\,G_D-\text{cost of the inserted transaction},\]
so \(\varepsilon b\,G_D\) exceeds the cost: \(\varepsilon b\,G_D>(\nu-\varepsilon)b\hat g\) for a provider and \(\varepsilon b\,G_D>\hat c+b\hat g\) for a consumer bid, which are the two bounds. \(\square\)

The providers of \(D\) are ones the honest crossing does not fund, which is why the builder needs the insertion to reach them. With \(\nu=1/2\), \(\beta=0.1\), \(\varepsilon=0.01\) and \(\hat g\ge g_0\), \(D\) must exceed 1.03 million gas for a provider and 2.31 million for a consumer bid. A seed helps the builder only with a fee cap near \(b\). With fee cap \(0.99b\), it takes \(0.02b\) per gas from the pool, while the consumers its added allowance admits pay \(\lambda\rho\) per gas of seed, at density \(\rho\); the pool thus gains \((\lambda\rho-0.02b)\) per gas of seed. The seed costs the builder \(0.99b-0.01b=0.98b\) per gas and needs 98 times its gas in \(D\), about 2.06 million for a 21{,}000-gas seed. So a gain needs more than 1.03 million gas of genuine providers, and about 2.06 million for a seed with fee cap \(0.99b\), that are waiting among the block's candidates but are not funded in \(B^\ast\).

\textbf{Lemma 3 (removing a selected consumer).} \emph{Fix the allowance \(A_C\) and the included consumer candidates. If one selected consumer is removed from the candidates, the other selected consumers pay in total at least as much as before.}

\emph{Proof.} Run the selection of \S~3.1 with and without the removed consumer \(h\). Until some candidate is treated differently, the run without \(h\) has \(\hat g\) more room, so the first candidate \(x\) treated differently is selected only without \(h\). In the run with \(h\), \(x\) did not fit, so the consumers that run selects after \(x\) have total gas below \(g_x\), each at density at most \(\rho_x\). Those of them that the run without \(h\) leaves out therefore pay less than \(\rho_xg_x=c_x\), which \(x\) adds. If no candidate is treated differently, the others pay the same. \(\square\)

The lemma is about the total because a single consumer can lose its place. Take allowance 10 and three consumers with gas 4, 7 and 6, in density order. The rule selects the first, skips the second, which does not fit in the 6 left, and selects the third. Without the first, it selects the second, and the third no longer fits in the 3 left. The third drops out, but the second pays at least as much per gas, on more gas. The proofs below need only the total, since the receipts \(R\) are what fund the providers.

Proposition~6 below is the main result of this section: it gives a condition on the link ratio \(\lambda\) under which no inserted transaction gives the builder a positive gain, and Corollary~1 extends it to any finite set of insertions. By Proposition~5, a gain needs more than 1.03 million gas of providers in \(D\). With \(\lambda\) large enough, those providers alone fill the consumer allowance, so the insertion is pure cost, and the builder is worse off by inserting its own transactions.

\textbf{Proposition 6 (no gain from insertion).} \emph{Let \(\lambda\ge\frac{\varepsilon K_C}{(\nu-\varepsilon)g_0}\) and \(\lambda\ge\frac{2\varepsilon K_C}{g_0}\). Then no block \(I\) with an inserted provider or an inserted consumer bid pays the builder more than \(B^\ast\), its best block with no inserted transaction.}

\emph{Proof.} Recall that the consumer allowance is \(A_C=\min\{K_C,\lambda G_P\}\), where \(G_P\) is the funded provider gas (\S~3.1). The two conditions say \(K_C/\lambda\le\frac{\nu-\varepsilon}{\varepsilon}g_0\) and \(K_C/\lambda\le\frac{g_0}{2\varepsilon}\).

Let \(I\) be a block with an inserted transaction of gas \(\hat g\), and \(D\) as in Proposition~5; \(D\) holds only genuine providers. If \(I\) gains, Proposition~5 gives \(G_D>\frac{\nu-\varepsilon}{\varepsilon}\hat g\ge\frac{\nu-\varepsilon}{\varepsilon}g_0\ge K_C/\lambda\), so \(\lambda G_D>K_C\) and
\[A_C=\min\{K_C,\lambda G_P\}=K_C\]
with or without the inserted transaction: the providers of \(D\) alone hold the allowance at its cap. We show that removing the insertion then raises the builder's payoff, so \(I\) does not gain.

\emph{Inserted provider.} Its gas adds to \(G_P\) but cannot raise \(A_C\) above \(K_C\). Without it, \(A_C\) is still \(K_C\), so the same consumers are selected and the receipts \(R\) are unchanged, while the pool pays less. Removing it saves the builder its net cost on that provider: as the sender it pays the fee cap \(m\) on \(\hat g\) gas, and as the builder it receives the inclusion fee \(p\), a net loss of \(\hat g(m-p)\ge(\nu-\varepsilon)b\hat g>0\) (Proposition~3).

\emph{Inserted consumer bid.} The bid adds no provider gas, so it does not change \(A_C\). It adds its payment \(\hat c\) to the receipts \(R\), and it costs the builder \(\hat c+b\hat g\): its declaration and the base fee it burns. When the bid is removed, the receipts fall by at most \(\hat c\), since the other selected consumers pay no less (Lemma~3), so the shortfalls \(S\) can exceed \(R\) by at most \(\hat c\). To restore \(S\le R\), the builder drops providers of \(D\) one at a time, smallest gas first, until the dropped shortfalls cover the excess; let \(x\) be the last one dropped. The providers dropped before \(x\) have shortfalls totalling less than \(\hat c\), since otherwise the dropping would stop before \(x\). A provider's inclusion fee \(g_jp_j\) is smaller than its shortfall \(g_js_j=g_j(b-m_j+p_j)\), because \(m_j<b\), so their inclusion fees also total less than \(\hat c\). Provider \(x\) earns at most \(g_0b\le b\hat g\), by the cap. The builder therefore loses less than \(\hat c+b\hat g\) in inclusion fees and saves \(\hat c+b\hat g\), the cost of the bid, so it earns more without the bid.

Dropping these providers keeps \(A_C=K_C\), because the providers of \(D\) that remain have at least \(K_C/\lambda\) gas. Some provider remains, since the fees on \(D\) exceed \(\hat c+b\hat g\), and each remaining one has at least \(g_x\) gas. If \(g_x\ge K_C/\lambda\), that is enough. Otherwise every dropped provider has less than \(K_C/\lambda\le g_0/\varepsilon\) gas and earns \(\varepsilon b\) per gas, so the gas dropped before \(x\) is less than \(\hat c/\varepsilon b\). Since \(\varepsilon b\,G_D>\hat c+b\hat g\), the remaining gas exceeds \(\hat g/\varepsilon-g_x\ge g_0/\varepsilon-g_x>g_0/\varepsilon-K_C/\lambda\ge K_C/\lambda\). \(\square\)

\textbf{Corollary 1 (several insertions).} \emph{Under the conditions of Proposition~6, no block \(I\) with a finite set \(H\) of inserted transactions pays the builder more than \(B^\ast\).}

\emph{Proof.} Suppose \(I\) pays more than \(B^\ast\), and let \(D\) be the providers, other than inserted ones, funded in \(I\) but not in \(B^\ast\). Every inserted transaction costs the builder a positive amount: an inserted provider at least \((\nu-\varepsilon)b\hat g\) (Proposition~3), an inserted consumer bid that is selected \(\hat c+b\hat g\), and one that is not selected the base fee \(b\hat g\). The argument of Proposition~5 gives \(\varepsilon b\,G_D\) above the total cost of \(H\), so \(G_D>\frac{\nu-\varepsilon}{\varepsilon}g_0\ge K_C/\lambda\), and \(A_C=K_C\) with or without the transactions of \(H\).

We show that removing the transactions of \(H\) gives a block that pays the builder more than \(I\). Taking out the inserted providers leaves \(A_C\) at \(K_C\), so the same consumers are selected and \(R\) is unchanged, while the pool pays less and the builder saves its net cost on each of them. Taking out an inserted consumer bid that is not selected changes nothing except that the builder no longer pays its base fee. Taking out the inserted consumer bids that are selected lowers \(R\) by at most \(\hat C\), the total of their declarations, since by Lemma~3, applied once for each, the other selected consumers pay no less. To restore \(S\le R\), the builder drops providers of \(D\), smallest gas first, as in the proof of Proposition~6. The inclusion fees it loses are below \(\hat C+g_0b\), which is at most \(\hat C+b\hat G\), the cost of those consumer bids with total gas \(\hat G\). The providers of \(D\) that remain keep \(A_C=K_C\), by the computation of Proposition~6 with \(\hat c\) and \(\hat g\) replaced by \(\hat C\) and \(\hat G\). The resulting block has no insertion and pays the builder more than \(I\), and so more than \(B^\ast\), which is impossible. \(\square\)

At \(\nu=1/2\), \(\varepsilon=0.01\), \(g_0=21{,}000\) and \(K_C=6\) million, the conditions read \(\lambda\ge5.83\) and \(\lambda\ge5.71\), so \(\lambda=6\) meets both; equivalently, the provider gas that opens the whole allowance, \(K_C/\lambda=1\) million, is below 1.03 and 1.05 million. They are needed: with \(\lambda=6\) and \(K_C=15\) million, a single waiting consumer of 14.53 million gas at the floor and providers of 1.2 million gas at shortfall \(0.51b\) do not cross, since two providers open too little allowance and three cost more than the consumer pays, but two providers and a 21{,}000-gas seed do, and the builder gains \(3{,}420b\). The link ratio also gives \(\lambda\beta=0.6\ge1-\nu+\varepsilon\), so a provider whose added allowance waiting consumers fill pays for itself at the consumer floor, and the honest crossing rarely leaves such providers waiting.

\subsection{What is not claimed}\label{not-claimed}

\begin{itemize}
\tightlist
\item \textbf{User incentive compatibility.} The consumer rule is first-price, so a consumer gains by declaring less than its value while it is still selected (\S~3.6). Truthful declaration is not claimed.
\item \textbf{Resistance to agreements outside the protocol,} the third property of {[}10{]}. The burned surplus is where such an agreement would gain, since payments that would be burned could instead be shared between a builder and users.
\item \textbf{Fair candidate selection.} Validity is checked relative to the proposed block, so a builder can omit an eligible candidate, as under EIP-1559 today. TLM adds a motive rather than a capability: a builder can favour an affiliate by omitting a higher-density consumer. Under pay-as-bid the omission does not change what the affiliate pays, and Proposition~3 still binds, so the gain is placement rather than extraction.
\end{itemize}

\section{Simulation study for TLM regulation under provider-gas bounds}\label{simulation-study-for-tlm-regulation-under-provider-gas-bounds}

This section evaluates TLM regulation with provider-gas allowances of 20\% and 30\% of the hard block limit. In both variants, funded provider gas is excluded from the ordinary EIP-1559 controller and separately bounded by the realized provider-gas allowance. Baseline EIP-1559 and na\"ive metering are comparison cases: na\"ive metering instead includes funded provider gas in the controller, which raises the base fee at a positive fixed point and reduces gas admitted through the ordinary channel.

The main mixture gives the following steady-state means. Parentheses contain 95\% confidence half-widths across 20 seeds.

Table 1. \textbf{Baseline} is unmodified EIP-1559. \textbf{N} is na\"ive metering: it applies TLA while metering provider gas against the ordinary EIP-1559 target. \textbf{T20} and \textbf{T30} apply TLM regulation with a TLA-funded provider-gas allowance of, respectively, 20\% and 30\% of the hard block limit.

{\def\LTcaptype{none} 
\begin{longtable}[]{@{}
  >{\raggedright\arraybackslash}p{(\linewidth - 12\tabcolsep) * \real{0.1111}}
  >{\raggedleft\arraybackslash}p{(\linewidth - 12\tabcolsep) * \real{0.1481}}
  >{\raggedleft\arraybackslash}p{(\linewidth - 12\tabcolsep) * \real{0.1481}}
  >{\raggedleft\arraybackslash}p{(\linewidth - 12\tabcolsep) * \real{0.1481}}
  >{\raggedleft\arraybackslash}p{(\linewidth - 12\tabcolsep) * \real{0.1481}}
  >{\raggedleft\arraybackslash}p{(\linewidth - 12\tabcolsep) * \real{0.1481}}
  >{\raggedleft\arraybackslash}p{(\linewidth - 12\tabcolsep) * \real{0.1481}}@{}}
\toprule
\begin{minipage}[b]{\linewidth}\raggedright
case
\end{minipage} & \begin{minipage}[b]{\linewidth}\raggedleft
base fee
\end{minipage} & \begin{minipage}[b]{\linewidth}\raggedleft
total utilization (\%)
\end{minipage} & \begin{minipage}[b]{\linewidth}\raggedleft
ordinary gas (million)
\end{minipage} & \begin{minipage}[b]{\linewidth}\raggedleft
funded provider gas (million)
\end{minipage} & \begin{minipage}[b]{\linewidth}\raggedleft
builder revenue (million)
\end{minipage} & \begin{minipage}[b]{\linewidth}\raggedleft
builder revenue change
\end{minipage} \\
\midrule
\endhead
\bottomrule
\endlastfoot
Baseline & 28.54 (0.05) & 50.05 (0.01) & 30.03 (0.009) & 0 & 49.89 (0.07) & baseline \\
N & 33.78 (0.05) & 50.07 (0.01) & 23.79 (0.036) & 6.25 (0.037) & 42.12 (0.08) & \(-\)15.57\% \\
T20 & 28.54 (0.05) & \textbf{69.33 (0.06)} & 30.03 (0.009) & 11.57 (0.032) & 53.19 (0.07) & +6.61\% \\
T30 & 28.54 (0.05) & \textbf{71.57 (0.15)} & 30.03 (0.009) & 12.91 (0.085) & 53.56 (0.07) & +7.37\% \\
\end{longtable}
}

The simulation implements the analytical contrast. T20 and T30 retain the baseline ordinary gas (30.03 million) and base fee (28.54), while raising total utilization from 50.05\% to 69.33\% and 71.57\% through, respectively, 11.57 and 12.91 million funded provider gas per block. N instead remains at 50.07\% utilization, raises the base fee 18.3\%, and reduces ordinary gas 20.8\%. Selected consumers pay 53.76 million per block. Under T20, 44.59 million of it funds providers and 9.17 million (17.1\%) is burned; under T30 the burn is 0.50 million (0.9\%). Two constraints therefore share the work in this mixture: consumer payments bound funding, and under T20 the provider-gas allowance binds in part of the blocks (realized provider gas averages 11.57 million against an allowance of 12.00 million, and 12.91 against 18.00 under T30). The allowance link does not bind in any block under T20, T30 or N, because funded provider gas exceeds \(K_C/\lambda=1\) million. The preserved ordinary gas and base fee follow from Lemma~2 rather than from measurement; the simulation adds the magnitude of funded provider gas and the incidence. Funded shortfall never exceeds consumer payments, and no reference-block transaction is removed. Appendix~A reports payment incidence and Appendix~B reports consumer-load sensitivity.

\textbf{Sensitivity to declarations.} The simulation draws declarations, and under a first-price rule consumers have reason to declare less than they would pay (\S~3.6). Table~2 varies the level of declarations under T20. The last row charges each selected consumer the least at which it is still selected when the others keep their declarations; it uses every fourth block, on which the first row gives 11.57 million.

Table 2. T20 in the main mixture, by level of consumer declarations: consumer payments per block, funded provider gas per block (both in millions), and total utilization. Ordinary gas is 30.03 million and the base fee 28.54 in every row.

\begin{center}\small\begin{tabular}{@{}l@{\quad}r@{\quad}r@{\quad}r@{}}
\toprule
declarations & paid (M) & provider gas (M) & utilization (\%) \\
\midrule
declarations as drawn & 53.76 & 11.57 & 69.33 \\
every declaration 20\% lower & 43.01 & 10.90 & 68.22 \\
every declaration 40\% lower & 32.04 & 9.53 & 65.93 \\
least charge at which still selected & 36.70 & 9.99 & 66.7 \\
\bottomrule
\end{tabular}\end{center}

Total utilization is 69.3\% at the drawn declarations and about 66.7\% when consumers pay the least that keeps them selected. The structural results do not depend on the level of declarations; the utilization figure does, and we report it as a range.

\textbf{Generalized second price, inserted bids, and the provider rules.} On the same blocks, a generalized second-price rule that charges each selected consumer the next density in the order funds 11.47 million provider gas at the drawn declarations, against 11.57 million under pay-as-bid, and 9.99 million at least charges, the same as pay-as-bid. Under it, 94\% of selected consumers gain by declaring less, by 30\% of the declaration on average.

We also test inserted consumer bids. Under pay-as-bid the builder may fund any unfunded provider and insert the cheapest consumer bid that completes its funding; a gain is counted net of that bid and of the base fee it burns. No insertion gains in 2{,}000 sampled blocks of the main mixture under T20, nor in 2{,}000 blocks under T30 or in the 10\% and 33\% consumer mixtures. These runs are numerical checks under the synthetic generator, with one insertion per block; none gains. The results rest on Propositions~4 and~6 and Corollary~1. We also insert seed providers of 21{,}000 to 1 million gas with fee caps from \(0.5b\) to \(0.99b\) in mixtures with 3\% to 20\% of offered gas from providers, where the link binds. At \(\lambda=1\) it binds in up to every block, but the unfunded providers that fit total at most 0.07 million gas, their deficit per gas is \(0.06b\) or more, and no seed gains. At \(\lambda=6\) it binds in at most 16\% of blocks, no unfunded provider fits, and no seed gains. Under the generalized second-price rule a search over bids placed just under each selected consumer, which is not exhaustive, finds no gain.

\textbf{The inclusion fee.} The fee sets the builder's revenue from the crossing and how far consumer payments go. At \(\varepsilon=0.01\), \(0.02\) and \(0.05\), T20 funds 11.57, 11.43 and 10.98 million provider gas per block, total utilization is 69.3\%, 69.1\% and 68.4\%, and builder revenue is 6.6\%, 13.0\% and 25.8\% above the baseline. No inserted consumer bid gains at any of these values, nor at \(\varepsilon=0.2\). A larger fee pays the builder more and funds less provider gas, and through \(\lambda\beta\ge1-\nu+\varepsilon\) it also needs a larger link ratio, so the fee is best kept at the smallest value that makes funding providers worth the builder's effort.

T20 and T30 cap provider gas at, respectively, 20\% and 30\% of the hard block limit. These are model parameters, not proposed Ethereum constants. The allowance bounds provider gas, not total occupancy: realized occupancy extends past \(T+\bar P_t\) because ordinary gas varies block to block. Against a baseline whose ninetieth percentile is 58.9\%, T20 reaches 78.6\%. Occupancy remains bounded above by construction: admission is bounded by \(K-H_t^0\), so no block exceeds the hard limit, and node requirements are provisioned against \(K\) rather than against realized occupancy. Raising utilization is an objective of TLM regulation and not a side effect. Providers are admitted outside the ordinary controller's target, so ordinary admission and the ordinary base fee are unchanged, and the added gas is demand the baseline suppressed rather than demand displaced from elsewhere. The effect of higher sustained utilization on node load and state growth is left to future work.

\section{Limits and research boundary}\label{limits-and-research-boundary}

\textbf{Simulation scope.} The reported single-slot study uses synthetic traffic to evaluate the provider allowance and base-fee response. Complementary work studies multi-slot TSP mechanisms and feedback-based congestion control; those results are not reported here.

\textbf{Consumer bidding.} The settlement rule is first-price and the simulation draws declarations. An equilibrium of declarations, with unequal realized gas, rebidding, and the allowance link, is left to future work; Table~2 brackets its effect on funded gas.

\textbf{Order-dependent gas.} Density and the allowance use realized gas, and a transaction's realized gas can depend on what executes before it. The builder must propose a block at which the checks pass; how often reordering changes the selected set remains to be measured.

\textbf{Deployment.} The block-local mechanism of \S~3 requires protocol support because it changes valid inclusion and settlement for providers with \(m_i<b\), and because a verifiable consumer early-region guarantee requires an ordering rule. This paper specifies the mechanism-level rule. A provider's tip field is fixed at zero (\S~3.3); how the TLA itself is carried, for example in a new transaction type, is an open question for the encoding. A follow-up will develop an EIP-ready transaction encoding and deployment path.

Future work extends the TLM program to welfare, fairness, strategic behavior, and anti-collusion; feedback-based congestion control; cross-slot TSP mechanisms; and applications to L2 and other L1 chains.

\section{Conclusion}\label{conclusion}

We introduce \emph{temporal liquidity}: the economically relevant capacity to move execution across admissible positions, together with the value that movement gains or surrenders. A transaction's position within a block and its execution delay can change its outcome and value. A temporal liquidity market lets users price that difference. A consumer can authorize payment for an early region and for its relative rank among selected consumers, while a provider accepts execution outside that region. Their crossing turns complementary temporal requirements into a self-funded mechanism rather than a one-sided fee bid, adding a time axis to dynamic fee-market design and congestion control.

We then apply this framework to Ethereum through EIP-1559. The block-local extension gives selected consumers early-region assurance and relative rank among selected consumers within it, and can fund below-floor providers that accept exclusion from that region. Our analysis shows why funded provider gas should be excluded from the ordinary EIP-1559 controller: counting it raises the base fee and displaces ordinary demand. Under TLM regulation, a bounded, consumer-funded provider allowance retains the reference ordinary allocation and raises total utilization in the modeled setting. Against strategic senders and builders, the rules are checked by every validator, the base fee is burned in full, no subsidy enters, and with a capped inclusion fee and a suitable link ratio no finite set of transactions a builder inserts raises its direct fee payoff.

\begin{credits}
\subsubsection{\ackname} The author thanks his former colleague at Amber Group, Adam Wang, who sparked his interest in blockchain research and generously shared his knowledge of this field. He also thanks Fan Zhang of Yale University, Chenyuan Wu and Wen Yan of City University of Hong Kong, Yijun Zhao of Fordham University, Norman Saade of the PBS Foundation, and a researcher at the PBS Foundation for helpful discussions and comments. Any errors are the author's own.
\end{credits}

\appendix

\section{Payment incidence in the main synthetic mixture}\label{payment-incidence-in-the-20-percent-consumer-synthetic-mixture}

This appendix reports secondary incidence results omitted from the main comparison. Columns Baseline, N, T20, and T30 denote unmodified EIP-1559, na\"ive metering, and TLM regulation with TLA-funded provider-gas allowances of 20\% and 30\% of the hard block limit. User entries are total class payments per block, including consumer TLA charges, of which the surplus over funded shortfalls is burned, and provider sender payments; builder entries are revenue. Parentheses show percentage change from Baseline. A dagger marks provider totals whose class composition differs from Baseline.

{\def\LTcaptype{none} 
\begin{longtable}[]{@{}
  >{\raggedright\arraybackslash}p{(\linewidth - 8\tabcolsep) * \real{0.1579}}
  >{\raggedleft\arraybackslash}p{(\linewidth - 8\tabcolsep) * \real{0.2105}}
  >{\raggedleft\arraybackslash}p{(\linewidth - 8\tabcolsep) * \real{0.2105}}
  >{\raggedleft\arraybackslash}p{(\linewidth - 8\tabcolsep) * \real{0.2105}}
  >{\raggedleft\arraybackslash}p{(\linewidth - 8\tabcolsep) * \real{0.2105}}@{}}
\toprule
\begin{minipage}[b]{\linewidth}\raggedright
party
\end{minipage} & \begin{minipage}[b]{\linewidth}\raggedleft
Baseline
\end{minipage} & \begin{minipage}[b]{\linewidth}\raggedleft
N
\end{minipage} & \begin{minipage}[b]{\linewidth}\raggedleft
T20
\end{minipage} & \begin{minipage}[b]{\linewidth}\raggedleft
T30
\end{minipage} \\
\midrule
\endhead
\bottomrule
\endlastfoot
neutral & 630.28m & 589.92m\break(\(-\)6.40\%) & 630.28m\break(0.00\%) & 630.28m\break(0.00\%) \\
consumers & 270.47m & 301.68m\break(+11.54\%) & 324.23m\break(+19.88\%) & 324.23m\break(+19.88\%) \\
providers\(^{\dagger}\) & 5.19m & 164.46m\(^{\dagger}\) & 293.83m\(^{\dagger}\) & 323.20m\(^{\dagger}\) \\
builder & 49.89m & 42.12m\break(\(-\)15.57\%) & 53.19m\break(+6.61\%) & 53.56m\break(+7.37\%) \\
\end{longtable}
}

Gas-weighted user payment gives a complementary view:

{\def\LTcaptype{none} 
\begin{longtable}[]{@{}
  >{\raggedright\arraybackslash}p{(\linewidth - 8\tabcolsep) * \real{0.1579}}
  >{\raggedleft\arraybackslash}p{(\linewidth - 8\tabcolsep) * \real{0.2105}}
  >{\raggedleft\arraybackslash}p{(\linewidth - 8\tabcolsep) * \real{0.2105}}
  >{\raggedleft\arraybackslash}p{(\linewidth - 8\tabcolsep) * \real{0.2105}}
  >{\raggedleft\arraybackslash}p{(\linewidth - 8\tabcolsep) * \real{0.2105}}@{}}
\toprule
\begin{minipage}[b]{\linewidth}\raggedright
user class
\end{minipage} & \begin{minipage}[b]{\linewidth}\raggedleft
Baseline
\end{minipage} & \begin{minipage}[b]{\linewidth}\raggedleft
N
\end{minipage} & \begin{minipage}[b]{\linewidth}\raggedleft
T20
\end{minipage} & \begin{minipage}[b]{\linewidth}\raggedleft
T30
\end{minipage} \\
\midrule
\endhead
\bottomrule
\endlastfoot
neutral & 30.18 & 35.43\break(+17.38\%) & 30.18\break(0.00\%) & 30.18\break(0.00\%) \\
consumers incl.~TLA & 30.19 & 42.26\break(+39.99\%) & 36.19\break(+19.88\%) & 36.19\break(+19.88\%) \\
providers\(^{\dagger}\) & 27.68 & 26.31\(^{\dagger}\) & 24.99\(^{\dagger}\) & 24.68\(^{\dagger}\) \\
\end{longtable}
}

\(^{\dagger}\) Provider totals are not percentage-comparable across cases: baseline provider-class gas contains only transactions whose fee caps clear the base fee, whereas TLA admits additional below-floor providers. Consumer entries divide the sum of the ordinary per-gas charge and the realized lump-sum TLA settlement by realized gas; the temporal component is not a per-gas price (\S~3.1). Neither table is a welfare estimate.

\section{Consumer-load sensitivity}\label{appendix-b.-consumer-load-sensitivity}

With 10\%, 20\%, and 33\% offered consumer gas, the realized gas of selected consumers averages 7.21\%, 9.77\%, and 9.92\% of \(K\), respectively. Unselected consumer gas averages 0.17, 3.10, and 8.99 million gas: the allowance \(K_C=0.10K\) is slack in most blocks of the first mixture, active in the second, and heavily oversubscribed in the third, where selected gas stays below \(K_C\) only because transactions are indivisible. T20 total utilization is 66.19\%, 69.33\%, and 69.78\%, with funded provider gas of 9.68, 11.57, and 11.84 million gas; T30 gives 66.68\%, 71.57\%, and 73.21\% with 9.97, 12.91, and 13.89 million. The 20\% early region is a completion boundary, not consumer-reserved capacity.

\section*{References}\label{references}

\sloppy
\begin{enumerate}
\def\labelenumi{\arabic{enumi}.}
\item
  Babaioff, M., Nisan, N.: On the welfare of EIP-1559 with patient bidders. In: EC '25. arXiv:2502.20031 (2025)
\item
  Buterin, V., Conner, E., Dudley, R., Slipper, M., Norden, I., Bakhta, A.: EIP-1559: Fee market change for ETH 1.0 chain. Ethereum Improvement Proposals (2019)
\item
  Capponi, A., Zhu, B.: Auctioning time to mitigate latency races: theory and evidence from blockchains. SSRN (2026)
\item
  Lavee, N., Nisan, N., Pai, M., Resnick, M.: Does your blockchain need multidimensional transaction fees? arXiv:2504.15438 (2025)
\item
  Lazzaretti, A., Papamanthou, C., Hishon-Rezaizadeh, I.: Robust double auctions for resource allocation. In: FC '25, LNCS 15751 (2025)
\item
  Leonardos, S., Monnot, B., Reijsbergen, D., Skoulakis, S., Piliouras, G.: Dynamical analysis of the EIP-1559 Ethereum fee market. In: AFT '21. arXiv:2102.10567 (2021)
\item
  Liu, Y., Lu, Y., Nayak, K., Zhang, F., Zhang, L., Zhao, Y.: Empirical analysis of EIP-1559: transaction fees, waiting time, and consensus security. In: CCS '22, pp.~2099--2113 (2022)
\item
  Penna, P., Schneider, M.: Serial monopoly on blockchains with quasi-patient users. In: FC '25. arXiv:2405.17334 (2025)
\item
  Reijsbergen, D., Sridhar, S., Monnot, B., Leonardos, S., Skoulakis, S., Piliouras, G.: Transaction fees on a honeymoon: Ethereum's EIP-1559 one month later. arXiv:2110.04753 (2021)
\item
  Roughgarden, T.: Transaction fee mechanism design. J. ACM (2024)
\item
  Zhao, Y.: The cost of delay: evidence from the Ethereum transaction fee market. SSRN working paper 4436697
\item
  Messias, J., Torres, C.F.: The express lane to spam and centralization: an empirical analysis of Arbitrum's Timeboost. arXiv:2509.22143 (2025)
\item
  DeFiLlama: Chain TVL, DEX volume, and active-address dashboards for Ethereum, Solana, and Tron. Snapshot collected 28 August 2026.
\item
  Artemis: Chain Compare, daily active-address series for Ethereum, Solana, and Tron. One-year series ending 25 August 2026.
\item
  RWA.xyz: Global Market Overview. Distributed tokenized-asset value and network shares, snapshot dated 2 April 2026.
\item
  Hyperliquid: HyperCore overview and order-book documentation. Fully onchain spot and perpetual order books and consensus-integrated ordering.
\item
  Kiayias, A., Koutsoupias, E., Lazos, P., Panagiotakos, G.: Tiered mechanisms for blockchain transaction fees. In: Leonardos, S., Alfieri, E., Knottenbelt, W.J., Pardalos, P.M. (eds.) \emph{Mathematical Research for Blockchain Economy: 5th International Conference, MARBLE 2024}, pp.~1--26. Springer (2024).
\item
  Franco, M., Rogozinski, G.: Mini-blocks: SSV-backed sub-slot auctions for Ethereum PBS. Ethereum Research (18 May 2026).
\item
  Lyu, X., Zhang, M., Zhang, X., Niu, J., Zhang, Y., Lin, Z.: An empirical study on Ethereum private transactions and the security implications. CoRR abs/2208.02858 (2022).
\item
  Arbitrum: Constitutional AIP: transition Arbitrum One ordering policy to priority gas auctions. Arbitrum Governance Forum (2026).
\item
  Botha, Q., Kyriacou, D., McMenamin, C., Oshitani, L., Stouka, A.-P.: SoK: preconfirmations. In: 2026 IEEE International Conference on Blockchain and Cryptocurrency (ICBC), pp.~1--19. IEEE (2026).
\item
  Ausubel, L.M.: An efficient ascending-bid auction for multiple objects. American Economic Review 94(5), 1452--1475 (2004).
\item
  Moulin, H.: Serial cost-sharing of excludable public goods. Review of Economic Studies 61(2), 305--325 (1994).
\item
  Paradigm: Priority is all you need. \url{https://www.paradigm.xyz/2024/06/priority-is-all-you-need} (2024).
\item
  Chung, H., Shi, E.: Foundations of transaction fee mechanism design. In: SODA '23. arXiv:2111.03151 (2023)
\item
  Edelman, B., Ostrovsky, M.: Strategic bidder behavior in sponsored search auctions. Decision Support Systems 43(1) (2007)
\end{enumerate}

\end{document}